# Huddler: Convening Stable and Familiar Crowd Teams Despite Unpredictable Availability

Niloufar Salehi, Andrew McCabe, Melissa Valentine, Michael Bernstein
Stanford University
{niloufar, msb}@cs.stanford.edu, {amccabe, mav}@stanford.edu

**ABSTRACT**
Distributed, parallel crowd workers can accomplish simple tasks through workflows, but teams of collaborating crowd workers are necessary for complex goals. Unfortunately, a fundamental condition for effective teams — familiarity with other members — stands in contrast to crowd work's flexible, on-demand nature. We enable effective crowd teams with Huddler, a system for workers to assemble familiar teams even under unpredictable availability and strict time constraints. Huddler utilizes a dynamic programming algorithm to optimize for highly familiar teammates when individual availability is unknown. We first present a field experiment that demonstrates the value of familiarity for crowd teams: familiar crowd teams doubled the performance of ad-hoc (unfamiliar) teams on a collaborative task. We then report a two-week field deployment wherein Huddler enabled crowd workers to convene highly familiar teams in 18 minutes on average. This research advances the goal of supporting long-term, team-based collaborations without sacrificing the flexibility of crowd work.

**ACM Classification Keywords**
H.5.3 Group and Organization Interfaces: Collaborative computing, Computer supported cooperative work

## INTRODUCTION
Crowdsourcing achieves impressive goals today by distributing work among independent individuals [6], but its future success will require collaborative crowd teams. Existing crowdsourcing techniques execute complex work [28] via pre-structured microtask workflows [6, 34]. However, seminal research from the field of organizational design has established that structured workflows are fundamentally incompatible for complex work [3, 50]. This literature suggests that completing complex work under uncertainty requires team-based coordination: teams iteratively establish a course of action, execute it, and then reflect and revise it based on their progress [21, 50]. Recognizing this, complex and creative goals such as design prototyping have prompted systems that assemble ad-hoc teams from the crowd [36, 37, 46].

Unfortunately, the on-demand nature of crowdsourcing would seem to make team-based coordination infeasible. Successful team-based coordination requires that team members build *familiarity* by working together repeatedly over time [15, 26, 45]. Familiar teams outperform ad-hoc teams by building common ground, learning to coordinate, and utilizing each person's unique skills [33, 45]. To reap these benefits, teams must keep the same members over time. However, stable team membership is not a core characteristic of crowd work: crowd workers on platforms such as Amazon Mechanical Turk (AMT) are available at unpredictable times [38] and often engage with other tasks when a new opportunity arises. Ad-hoc crowd teams on AMT feature an ever-changing roster of members (e.g., [46]), making it infeasible to build familiarity, and inhibiting the crowd's ability to achieve complex work.

In this paper we present Huddler, a system that enables assembly of familiar crowd teams, even under unpredictable availability and strict time constraints. With Huddler, crowd workers align themselves with any number of teams and request the appropriate team when they accept a task. If a team member is unavailable, Huddler recruits an alternative who maximizes team familiarity, as measured via the number of tasks previously completed with current team members. Huddler's crowd teams thus maintain a stable core, and bring in familiar faces as peripheral replacements. Huddler must recruit these members under a strict time limit, with knowledge that many workers will not respond or will decline. The system thus modulates its recruitment by measuring how likely a worker is to respond within a given time limit. Planning who to ask, and how long to wait for them before moving on, is a combinatorial problem with an exponential number of possible alternatives. We introduce a dynamic programming algorithm that allows Huddler to compute an optimal recruitment plan under real-time performance constraints.

We first demonstrate that familiarity improves the performance of teams of crowd workers. This effect is known for organizational work teams, but the lack of face-to-face interaction and shared organizational context may make it harder for crowd teams to reap the same benefits of familiarity [44]. We thus performed a field experiment in which 96 workers from Amazon Mechanical Turk (AMT) worked in teams of 3–4 across a series of 5 tasks to create Google advertisements for Kickstarter projects. We randomly assigned workers to either the *familiar* condition, where workers stayed with the same team through all 5 tasks, or the *ad-hoc* condition, where we re-randomized them into different teams in each round. The fifth and final advertisements were hosted publicly on Google AdWords. Familiar teams' advertisements received twice as many clicks as ad-hoc teams' advertisements ($p < .01$).

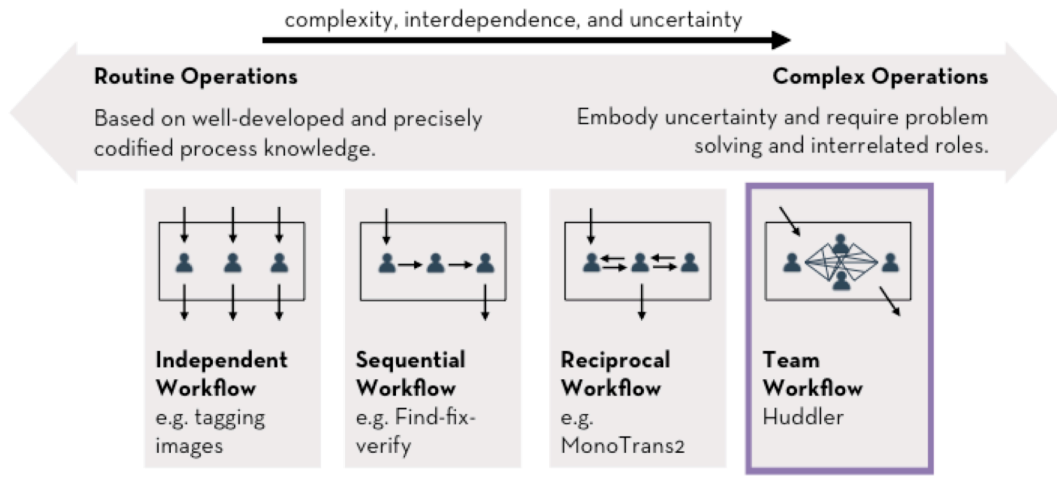

Figure 1. The Process Knowledge Spectrum characterizes work based on the maturity of our knowledge about how to produce a desired result [13]. At one end, work relies heavily on codified processes; at the other, work requires team-based collaboration. The appropriate workflow for a goal depends on the complexity of that goal [50]. In this paper we focus on enabling team-based crowd work toward the end-goal of achieving extremely complex operations.

Armed with the knowledge that familiar crowd teams achieve performance benefits, we deployed Huddler and evaluated its performance by recruiting 211 workers from AMT onto the system and posting tasks to the system over seven consecutive workdays. A 2x2 between-subjects study design modulated Huddler's recruitment algorithm by two factors: 1) *familiarity*, whether Huddler maximized familiarity when recruiting or treated all alternatives equally, and 2) *availability*, whether Huddler personalized its estimates of response time when recruiting or treated all alternatives as having equivalent response times. Huddler must balance the tradeoff between taking whoever is most likely available at that time against waiting longer to form more familiar teams: here we test the implications of each factor. The full system ("Huddler": high familiarity, high availability) successfully recruited teams for all tasks within the time limit, achieved the highest average pairwise familiarity between team members, and recruited teams equally quickly as the condition that considered only availability. By the final day of our experiment, teams formed in the Huddler condition had 10x higher familiarity scores than the control condition, and took on average 18 minutes to form, while teams in the control condition waited for over an hour. The system evolved this strategy throughout the study deployment: by balancing familiarity with availability, Huddler increased familiarity among people who were often also available at the same time.

This paper introduces the goal of designing for persistent and bounded teams in crowd work, and contributes technical solutions to the challenge of availability for such teams. We use AMT as our lever to demonstrate the benefits of familiar crowd teams. Unpredictable availability limits most crowd work platforms, therefore these concepts aim to apply beyond microtasking and to groups of crowd experts as well [46, 20, 31].

## RELATED WORK
In this paper, we draw on organizational behavior literature studying effective teaming in organizations. We also extend prior work exploring team based crowd work and supporting ad-hoc and virtual teams.

### Process knowledge: routine vs. complex operations
Process knowledge is defined as the knowledge of how to produce a desired result. The Process Knowledge Spectrum (Figure 1) characterizes work based on the maturity of our knowledge about how to achieve a goal [13]. When process knowledge is mature, for example in an assembly line, employees follow a prescribed set of instructions and get a certain result. Within such *routine operations*, uncertainty is minimal. On the other end of the spectrum lie domains in which we have less process knowledge, termed *complex operations*. Complex operations are more challenging. In these settings workers must form teams with complementing skills and interact continuously to create and execute a course of action [21].

Prior work on crowdsourcing has largely focused on supporting operations through independent, sequential or reciprocal workflows, which are more suited to routine operations (Figure 1). Pre-defined workflows mediated by algorithms guide crowd workers to complete bounded tasks that have a pre-defined input and output and are designed to be independent of the individual completing them [6]. Often, crowd workers have no knowledge of the global view or of the final goal. This approach ranges from applications that summarize text [4] and assist the blind [5] to expert oriented tasks as in flash teams [46]. The process knowledge spectrum demonstrates that as operations become more complex, interdependent, and uncertain, structured workflows struggle to support workers. Therefore, we explore the opportunity of using team workflows (Figure 1) as a more suitable mechanism to support crowd workers in achieving complex goals. Prior research has shown evidence of interactive groups of crowd workers outperforming individuals [54].

**Familiarity in teams**
Hackman describes teams as bounded, persistent groups of people who work together towards a common goal [21]. In this definition, teams are stable over time and members of a team gain familiarity with one another. As a result, they become more effective at group coordination over time [50]. Teams are also bounded, meaning that it is explicitly clear who is on the team and who is not, and all members share responsibility in the team's final outcome. This creates grounding for team members to cooperate, utilize each member's unique skills, and mentor one another [42].

Familiarity is an important factor that enhances team performance [26]. Prior work has found performance benefits of familiar teams both in controlled experiments [18, 22] and in long-term real world organizations such as software firms [26]. By working together, teams acquire situated knowledge that promotes more effective teamwork [45]. Over time, members discover each other's strengths and weaknesses, enabling them to act as a coordinated whole [21]. Working together also creates opportunities to create team beliefs such as psychological safety that in turn may increase knowledge sharing [14, 25]. Building on this literature, we focus on team familiarity as an important, yet understudied aspect of crowd teams.

**Computational systems, team formation, and team functioning**
The above research on complex work, team workflows, and team familiarity relates to teaming in organizations. A separate body of research focuses on how computational systems have enabled team formation and functioning among distributed and ad-hoc teams.

*Team formation*
Computational systems have allowed for new kinds of team formation by enabling open-calls to online crowdsourcing markets and volunteer crowds. This new way of forming teams means that crowd teams have emerged in a variety of online and unbounded settings. For example, crowd groups were able to translate Spanish poetry to English [29], teams of microtask workers were able to quickly translate an interface sketch into structured user interface elements [32], and flash teams coordinated expert crowd workers to complete tasks such as design prototyping and film animation [46]. Crowds of volunteers have collaborated to create Wikipedia articles [27] and create situational awareness during natural disasters [49]. Crowds have also collaborated in ways that were not planned in advance by system designers. The MIT team at the DARPA Red Balloon challenge won by coordinating an ad-hoc effort [52], and crowd workers on AMT have created ways to collaborate behind the scenes in their everyday work [17].

These crowd teams were formed using computational systems, but in a relatively emergent fashion. Researchers have also begun to make progress on how to use computational systems for more strategic or deliberate team formation. For example, researchers have designed systems that enable ad-hoc team formation within an organization based on different members' competence and expertise. When a new customer request arrives, the system forms an ad-hoc team that covers all areas of expertise needed for the task [23]. Additionally, ad-hoc groups in crowd work have been shown to have higher performance when they are formed with balanced personality types [36], which suggests that demographics or personality types could be algorithmically optimized in team formation.

Thus, computational systems have enabled new team forms and have begun to support deliberate and strategic team formation. In forming crowd teams there is a fundamental tension between complex work, which requires familiarity, and on-demand crowd work, which depends on workers' availability. Huddler addresses this issue by tracking connections between crowd workers and then utilizing this data to form relatively stable teams despite unpredictable availability. As a result, these on-demand teams can reap many of the benefits of long-term team membership.

*Team functioning*
Computational systems have also been designed to enable more effective collaboration among distributed or ad-hoc teams. These solutions are important because distributed collaboration has been shown to be extremely challenging [24, 41]. Anonymous online groups have trouble achieving trust among members and coordinating effectively [48], and online discussions may easily end in flames [10].

Distributed groups are most likely to succeed when they have high common ground and when the nature of their work is loosely coupled [44]. For groups with low common ground, as in crowdsourcing groups, structured workflows offer one possibility to make work more independent and decoupled. However, research on organizational design has shown that as goals become more complex, structured workflows become less applicable. Therefore, crowd teams face an uphill battle in performing effectively.

Researchers have studied how computer-mediated technology can support ad-hoc distributed groups that lack stable membership, norms, and routines. For example, email enables ad-hoc groups to exchange information more effectively, partition work more successfully, and increases participation [8]. Similarly, virtual co-presence facilitates trust among members and enhances performance in decision-making tasks [1]. Researchers have also designed tools for reflection in ad-hoc teams to catalyze group learning and increase shared understanding [43]. We use these lessons in the design of Huddler by providing teams with computer-mediated technology such as group synchronous chat and shared documents.

## STUDY: FAMILIARITY ENHANCES PERFORMANCE IN CROWD TEAMS

Familiarity enhances offline teams' performance [21, 45]. Does this result generalize to crowd teams? While working together should promote more effective coordination [45], distributed teams face significant coordination challenges due to the nature of their communication [44]. It is not clear whether, empirically, familiar crowd teams would provide benefit over current ad-hoc crowd teams. If familiar crowd teams do perform better, even at very short time scales, it becomes incumbent to design techniques that can reconvene the same teams repeatedly over time. Therefore, we conducted an experiment comparing workers' performance on a creative, collaborative task when they worked in *ad-hoc* vs. *familiar* teams.

### Task Choice

To measure team performance we needed a task that had the following properties:

- *Complex,* so that there is added benefit to working with a team and members have to coordinate their efforts
- *Not breakable into independent subtasks,* so that members have to collaborate
- *Include elements of creativity or uncertainty,* so that the task cannot have a pre-defined routine

We first piloted our experiment with items from the collective intelligence battery [51], for example creative brainstorming and solving visual problems. However, workers broke these tasks down among themselves and solved their own parts individually (e.g., splitting up forty puzzles up, ten per person), making them less useful at testing team effectiveness at interdependent tasks.

Therefore we chose a collaborative task that could not be broken down into individual subtasks: creating short advertisements. Prior work has also used this task to assess creative outcomes [11]. Specifically, we asked teams to review a Kickstarter product and create a short web advertisement for it. Following the requirements for Google advertisements, the ads consisted of a title of up to 25 characters, and two lines of description, each up to 35 characters. Teams used a collaborative text editor with an embedded chat client to perform the task together. Advertisements are not as complex as many team-based activities, for example software development. However, they are much more complex and creative than the work typically performed on Mechanical Turk [55], they are challenging to execute well, and they represent a multi-billion dollar industry. In practice, we found that the ads' short length limit was an advantage to our task. The tight constraints forced participants to brainstorm, discuss, and choose the best way to present the product.

### Method

We recruited 200 workers from Amazon Mechanical Turk (AMT) and randomly assigned each worker to either the ad-hoc or familiar condition. Our tasks were open to workers who: Lived in Canada or the US, had successfully completed at least 1000 tasks on the platform, and had an approval rating of 99% or higher. We used these qualifications to ensure that our crowd workers were fluent in English and would produce high quality results. We compensated workers according to the Dynamo guidelines for academic requesters[1] that require payment of at least the federal minimum wage in the US ($7.25/h at the time of writing). We paid $10 per task and each task took at most 1 hour to complete. We also offered a bonus of up to $2 for above average work.

Following methodology from prior literature [37, 39], we placed workers in a staging area to wait for up to 5 minutes while other workers joined. When enough people had joined the staging area, we divided them into teams of three to four people who shared the same study condition. Therefore, all team members began unfamiliar with each other. There is an inverse relationship between the number of people in a group and individual performance, so smaller group sizes are preferred [30]. Prior research has followed this principle to limit crowd teams to five people [36]. Due to the short time limit on our tasks we decided to minimize coordination costs by limiting teams to size three or four.

Each team had 10 minutes to complete each advertisement, after which we automatically saved their work and redirected them back to the staging area. While in the staging area, teams were re-shuffled if necessary. Teams repeated this process across five different products.

In the ad-hoc condition, teams were randomly re-shuffled between each task. In the familiar condition, teams were kept together unless they fell below a minimum size threshold due to dropouts, in which case they were recombined with other teams or dropped if none were available. Thus, workers in the ad-hoc condition would likely collaborate with new team members for each task, and teams in the familiar condition would likely work with the same team members across all tasks.

As an objective measure of team performance, we posted all of the resulting ads on Google Adwords and measured their clickthrough rate (CTR) — the percentage of advertisement impressions that resulted in a click [11]. Baseline clickthrough rates differ between products based on the popularity of that product and the keywords associated with it. Therefore, we compare CTRs within products, rather than between products. To compare team performance between conditions we performed an unpaired t-test on the CTRs for ads created at each step.

---

[1] http://guidelines.wearedynamo.org

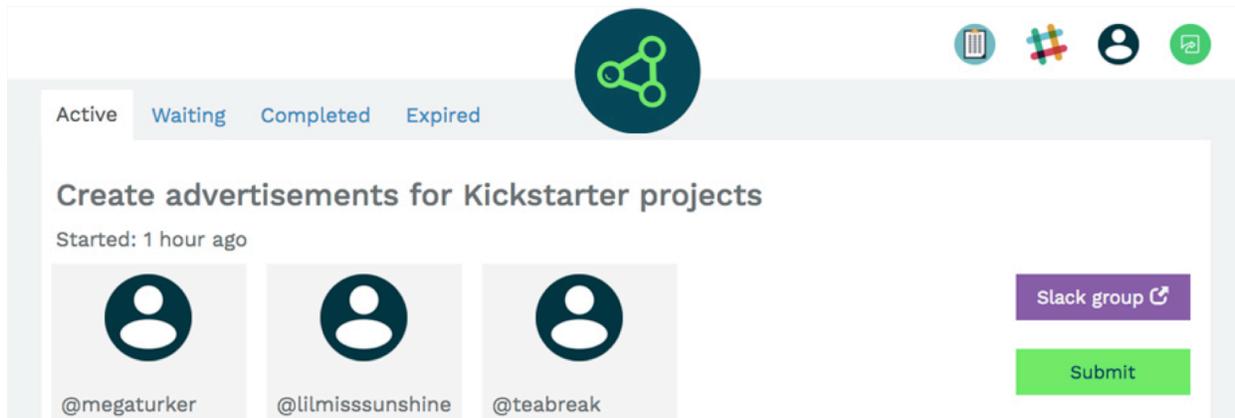

Figure 3. Huddler supports crowd workers to form persistent teams and work on tasks together.

**Results**

Teams submitted a total of 193 ads for 5 different Kickstarter projects. Clickthrough rates for advertisements submitted in the 1st through 3rd tasks showed no significant difference between the two conditions (Figure 2, all $p > .05$). However, by the fourth task, advertisements by teams in the familiar condition had significantly higher CTRs ($M = 0.9, SD = 0.5$) than those in the ad-hoc condition ($M = 0.5, SD = 0.4$), $t(33) = 2.30, p < .05, d = 0.7$. In the fifth task, the difference grew and advertisements in the familiar condition had twice the CTR ($M = 0.3, SD = 0.1$) as those in the ad-hoc condition ($M = 0.14, SD = 0.1$), $t(31) = 3.37, p < .01, d = 1.2$. Note that the absolute value of the CTR for each task is dependent on the popularity of that product and the keywords associated with it, therefore we can only compare CTRs within each task across the two conditions and not between different tasks.

At each step of our experiment a few workers dropped out, but 79% of dropouts happened during or right after the first task. In the second task, 40 teams submitted ads, 20 in each condition. In the fifth task, 33 teams submitted ads, 18 teams in the familiar condition and 15 teams in the ad-hoc condition. There were no significant differences in the number of dropouts in each condition (6 vs. 10), $X^2(1) = 1, p > .1$. A few workers contacted us later to say that they were forced to leave the task due to technical issues or to attend to a family member. This observation suggests that supporting familiar teams in a crowd work environment will require designing methods to handle unavailable team members.

These crowd teams worked together for only an hour yet the teams in the familiar condition had *twice* the performance of ad-hoc teams. Thus familiarity is an enabling condition for team performance even for teams of distributed crowd workers. In the design of crowd team systems, ensuring high familiarity is an important goal. We build on this knowledge to design Huddler.

**HUDDLER**

Huddler is a web platform that supports crowd workers in convening familiar teams for crowdsourcing tasks (Figure 3). When a team member is absent, Huddler automatically finds the most familiar alternative who is likely to respond within the given time frame and invites them to the team. In this section, we introduce Huddler and its techniques for managing crowd teams.

**Scenario**

Suppose a requester wanted to create a team task on Amazon Mechanical Turk. They specify team parameters such as team size and time limit on Huddler, and the system generates a unique "Find Team" button that the requester can copy-paste into the HTML of their task.

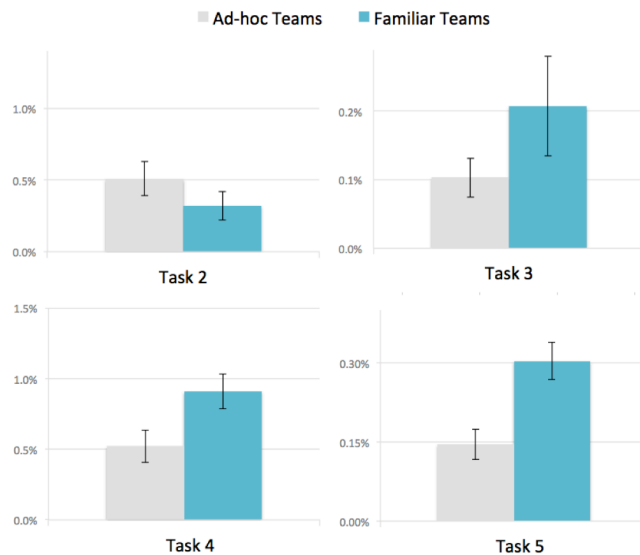

Figure 2. Clickthrough rates for teams in the *familiar* condition were twice as high as those in the *ad-hoc* condition by the final round of the study. CTR values differ across tasks due to product popularity and keywords. Task 1 did not receive any clicks.

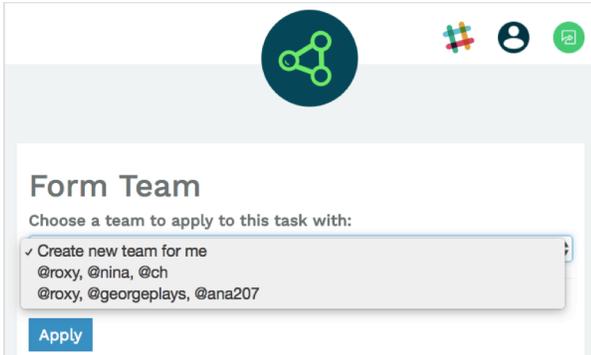

**Figure 4. A crowd worker can choose to apply to the task with a previous team or request a new one.**

Once the task is published, a crowd worker finds it in the marketplace and forms a team to complete it. They click on the "Find Team" button, and are redirected to the Huddler platform; here the task is reserved for them until it expires. Huddler asks them to pick a team — they can choose a previous team or request that Huddler find a new team for them (Figure 4). Workers can also author new teams with people that they would like to work with. The worker then waits for all team members to accept the invitation. During this period they can leave Huddler and spend time on other tasks. Huddler will automatically complete the team and send an email to all members when the task starts.

**Huddler hot-swaps missing members**
When a worker applies to a team task, Huddler sends invites to all team members. Each invitation has a time period for the worker to respond before it expires. If an invite expires or if the worker declines it, Huddler automatically hot-swaps that member with someone else and sends an invite to the new alternative member.

Our goal in the hot-swap process is to preserve the familiarity of team members as much as possible given the time limit before the task expires. In other words, Huddler must balance (1) workers who have a strong history of collaborations with existing members but who are unlikely to respond in the time allotted, against (2) workers who do not have a strong history of collaborations with these members but are very likely to be available.

There are two components to Huddler's assignment algorithm. The first is the *benefit* gained by adding a person to the current team. Benefit captures the added value that this person would bring to the team given its current members. The second is the member's *availability*, or the probability that they will respond within a given time frame. In this section, we provide a high-level overview of the hot-swap algorithm. Details follow in a separate section.

Given the strong effects of team familiarity in our experimental results, we model *benefit* as the new worker's past experience with current accepted team members. In settings where teams are stable over time, prior work has also measured familiarity as the cumulative experience of team members [9, 21]. For example, if a person $p$ were to join team $t$, $p$'s added benefit to this team would be the number of previous tasks that $p$ had completed with each other member. If $t$ already contained members $p_i$ and $p_j$, the benefit of adding $p$ to the team is the number of tasks that $p$ had completed with $p_i$ or $p_j$ previously on Huddler. So for $p$, having a history of prior collaborations with existing team members would make them a preferred candidate over others.

Huddler operates with any benefit function, and can be expanded to incorporate other factors. For instance, prior work has shown that crowd teams with balanced personality types perform better [36]. After gathering personality information from workers in the system, Huddler could prioritize new team members who would contribute to a balanced team. We can use the same strategy to support onboarding new users by temporarily prioritizing them in the benefit function. Other benefit features might include relevant expertise [46] or social perceptiveness [51].

Finally, Huddler must convene the team under a fixed time constraint. To do so, Huddler requires an algorithm that maximizes benefit under uncertainty about whether an invitee will accept and with the constraint of a strict time limit. Achieving this requires that Huddler model its uncertainty about whether an invited team member will accept. It does so by modeling each worker's probability of response based on prior observations in an empirical CDF (cumulative distribution function). For example, worker $p$ may have a .4 probability of responding within 30 minutes, a .6 probability of responding within one hour, and a .8 probability of responding within eight hours. To simplify our model, we only posted tasks in the morning and limited our pool to US workers to account for varying time zones.

Huddler maintains a record of every task invitation sent and the time passed until it was accepted, declined, or expired. Not every worker has a history of working with Huddler, so Huddler must bootstrap its model for new users. It does so by aggregating data from all users to generate a global empirical CDF. As Huddler gathers more data from each individual, it places more weight on that user's empirical CDF and less weight on the global CDF. Specifically, assuming Huddler has $n$ data points for a worker $w$, it calculates that worker's expected CDF:

$$CDF_{expected} = \frac{CDF_{global} + n * CDF_w}{(n+1)}$$

Huddler uses these measures of *benefit* and *availability* to choose the best new team member given the time constraint and invites them to the task. If their invitation expires or they decline, Huddler runs the same algorithm again to invite another member.

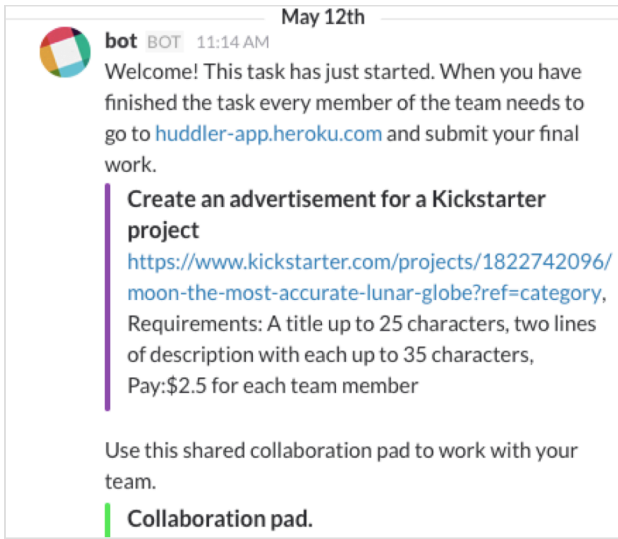

**Figure 5. Huddler automatically creates a channel in Slack and an instance of Etherpad for the team to collaborate.**

**Messaging and shared editors support collaboration**
Once the team is formed, Huddler sends a notification email to all members notifying them that the task has started and provides the team with support for collaboration. Huddler automatically creates a channel for the team in Slack[2] (Figure 5) for instant messaging and an instance of Etherpad for collaborative writing.

**Ratings inform future collaborations**
When the team has finished working, one member needs to submit the team's work. But the task isn't complete until all workers evaluate each other on a 3-point scale: (-1) "I did not like working with this person. Please don't pair me with them again!", (0) "Working with this person was fine but they're not my first choice.", and (1) "I really liked working with this person and want to work with them again!"

In the next section we will show how Huddler uses this information to calculate benefit scores.

**Assignment Algorithm**
In this section we explain Huddler's hot-swap assignment algorithm in more detail.

In inviting a new member to a team, Huddler calculates that person's benefit score based on their past collaborations with current team members. However, past collaborations are only beneficial if they were positive experiences. Therefore, in calculating pairwise familiarity scores between two people, Huddler not only calculates the number of past tasks that the two have completed together, but also weights each task by the mutual ratings that workers assigned to each other after the task.

---

[2] http://www.slack.com/

We define $PairwiseFamiliarity(p_i, p_j)$ as the number of tasks that $p_i$ and $p_j$ have worked on together weighted by their mutual ratings. For example, if after completing a task, $p_i$ and $p_j$ each rate the other with +1, their pairwise familiarity would increase by 2. But if $p_i$ rates $p_j$ with 0 and $p_j$ rates $p_i$ with -1, their pairwise familiarity would decrease by 1. Therefore, for each potential new team member $p$ and team $m$ we calculate a benefit score:

$$benefit(p, m) = \sum_{p_i \in m} PairwiseFamiliarity(p, p_i)$$

We seek to maximize this measure of benefit, while balancing it against estimated availability. Huddler models availability as $availability(p, t)$, the probability that $p$ will respond within the next $t$ hours if Huddler invites them now. For instance, inviting a person with very high benefit score will be useless if we don't expect them to respond within the given time limit.

To integrate benefit and availability, we use an expected value calculation. The expected value of inviting team member $p$ to a team $m$ under time limit of $t$ hours is the probability that they can join multiplied by their benefit score if they join:

$$E(p, m, t) = availability(p, t) \cdot benefit(p, m)$$

Therefore, for each person $p$, team $m$, and time limit $t$, we have established a measure of the expected benefit gained by inviting this person and waiting for them for $t$ hours.

For each task, Huddler's algorithm needs to take as input $T$, the time left until the task expires, and compute an ordering of people to invite, with time limits for each invitation. For example, Huddler might plan to invite high-familiarity worker $w_1$ and wait 30 minutes for them to respond, followed by inviting lower-familiarity worker $w_2$ and wait one hour, followed by asking a low-familiarity but extremely highly available worker $w_3$ for the last thirty minutes of the time budget just to ensure that the team is full in the worst case. Huddler must thus consider inviting all workers at all possible time points in order to find the plan that maximizes expected value. This is a combinatorial problem with an exponential number of possible plans.

However, the problem can be solved recursively. Furthermore, sub-tasks are repeatable: $E(p, m, t)$ takes on the same value always when the inputs are the same, because $benefit$ does not change and we assume a worker's $availability$ is fixed within the short time bounds of a Huddler recruitment phase. Thus, we take a dynamic programming approach that can solve this problem to find the optimal plan efficiently.

First, Huddler generates a list $L$ of potential workers to invite to the task and orders them by their $benefit$:

$$L = \{p_1, p_2, \ldots, p_n\}, \ benefit(p_i) > benefit(p_{i+1})$$

Huddler will go through the list in this order and invite members. Ordering people by $benefit$ produces an optimal solution because if one person accepts the invitation we no longer need to call the next people in the list. Next, Huddler needs to assign a time limit for waiting for each member. This time limit may be zero, in which case Huddler will pass over that member and invite the next person.

First we expand the definition of $E(p, m, t)$ to also accept as input a list of people of length $n$. Therefore, $E_l(L, m, t)$ is similarly defined as the expected value of optimally inviting the members of list $L$ in order over a time period of $t$. In practice we quantize $t$ to time periods of 30 minutes (e.g., 30min, 1hr, 1.5hr) to prevent crowd worker frustration with seemingly erratic response time demands. The resulting optimization problem is:

Maximize:

$$E_l(L, m, T) = \sum_{i=1}^{n} availability(p_i, t_i) \cdot benefit(p_i, m)$$

Subject to:

$$\sum_{i=1}^{n} t_i \leq t$$

Based on this definition, our dynamic programming algorithm relies on the following recursive solution:

$$E_l(L, m, T) = \sum_{t=0}^{T} E_l(L - p_1, m, T - t) + E(p_1, m, t)$$

Huddler recursively solves this problem, using dynamic programming lookup whenever it encounters previously computed subvalues, to produce an optimal solution. It then invites the first person on the list and assigns them the computed time period to respond to the invitation.

## EVALUATION

Huddler supports crowd workers to form teams and work on new tasks together with these familiar teams. Here, we report on a field experiment of Huddler that explored how effective Huddler was at supporting familiar crowd teams. We assessed how two experimental factors —familiarity and availability — impacted teams over a longer period, when member availability would fluctuate over time.

### Method

To evaluate Huddler, we recruited 211 workers from AMT to the system. We invited all workers who had completed our initial study. We also posted a qualification task on AMT and invited workers who completed it to Huddler. We used one of the advertisement tasks from our initial study as the qualification task so all workers on Huddler had basic familiarity with the task.

We sought to understand how each factor of Huddler's algorithm impacted the teams that were formed. Therefore we utilized a 2x2 between-subjects design modulating Huddler's algorithm by:

**Familiarity**: in the high familiarity condition, Huddler uses a member's history of collaborations with current team members to measure $benefit$ scores; as a result the algorithm maximizes familiarity between team members. In low familiarity, Huddler treats all members equally.

**Availability**: in the high availability condition, Huddler personalizes its estimates of response time for each member. In the low availability condition, it assumes the same response time distribution for everyone, which is based on aggregate responses from all users.

Based on this design, we randomly assigned each worker to one of four conditions: *Huddler*: high familiarity, high availability; *Familiarity*: high familiarity, low availability; *Availability*: low familiarity, high availability; *Control*: low familiarity, low availability. Therefore the condition that utilized the full system, *Huddler*, benefited from both the familiarity factor and the personalized availability measures. We use the *control* condition as a benchmark that does not differentiate between individual workers.

To ensure that people could not form teams that crossed experimental conditions, we temporarily disabled the option for workers to author new teams and removed teams that workers had already created in our piloting phase.

We posted 40-50 team tasks on Amazon Mechanical Turk (AMT) every morning over 7 consecutive weekdays. Each task had a time limit of 3 hours after which it would automatically expire. To complete a task, a crowd worker had to accept the task on AMT and then apply to the task on Huddler, where they would request that Huddler create a new team for them. Huddler formed teams of people within the same experimental condition. Workers in the high familiarity conditions (*Huddler* and *Familiarity*) were also given the option of choosing a previous team to work with.

Each worker could accept at most one task on AMT each day, meaning that they could form only one team per day. Workers could join additional teams each day if they were invited via Huddler. We used the same tasks as in our initial study: Workers were givenWe gave workers a random product from a set of 30 different Kickstarter products, and asked them to work with a team of three to create a short web advertisement for it. We conducted the experiment from over seven consecutive weekdays in May 2016.

We measured how many tasks each condition successfully completed, how long each team had to wait to form, how many invites were sent out before the team was formed, and the familiarity scores of the final teams.

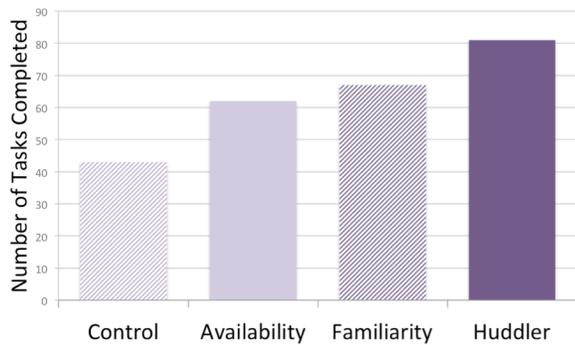

**Figure 6. Teams in the Huddler condition were most successful at forming teams and completed 1.8x more tasks than the control condition.**

### Results

Workers applied to 280 team tasks on Huddler. However S, some requests expired before a team was formed. Overall, teams on Huddler submitted 249 ads for 30 different products.

Familiarity and availability both increased the number of tasks that workers successfully completed (Figure 6). Workers in the Huddler condition completed 81 tasks, which was the highest among all conditions. In comparison, workers in the control condition completed 43.

To analyze this difference we performed a logistic regression analysis to predict the effect of each factor on completing a task. We found a positive effect of familiarity ($\beta=0.46, p<.01$) and of availability ($\beta=0.35, p<.05$).

The number of tasks that workers in a condition completed was a combination of how many tasks they accepted on AMT and how successful the system was at recruiting a team — which in turn affected their motivation to apply to a task later. For example, tasks in the control condition took more than an hour to start on average because the system invited an average of 2.6 unavailable workers to each task before finding available workers to form the team. Additionally, in this condition, 17% of tasks expired before all members had joined. Long waiting times and expired tasks made these workers less motivated to apply to our tasks later. However, the other three conditions did not have the same experience. Teams in the availability and familiarity condition each had one task expire, and no tasks expired within the Huddler condition.

*High familiarity resulted in increased familiarity scores for teams (Figure 7).* Teams in the Huddler condition had the highest familiarity scores, followed by teams in the familiarity condition. By definition, familiarity scores increase over time as teams complete more tasks together.

On day 1, as expected, there was no difference in the familiarity scores of teams formed in each condition. Comparing familiarity scores towards the final days may be misleading because teams in the Huddler condition completed many more tasks, and familiarity scores rise with more tasks completed. To account for this, we first compare familiarity scores midway through the experiment, when teams in each condition had still completed roughly the same number of tasks. On day 4, Huddler teams were the most familiar (M=31.4, SD=14.5), followed by the familiarity condition (M=13.1, SD=11.8), the availability condition (M=9.8, SD=9.2), and the control condition (M=5.6, SD=5.4). A two way ANOVA found a significant effect of familiarity ($F(1,32) = 13.5, p < .001$) and of availability ($F(1, 32) = 8.4$, $p < .01$), and no interaction effects ($F(1, 32) = 3.3$, $p > .05$).

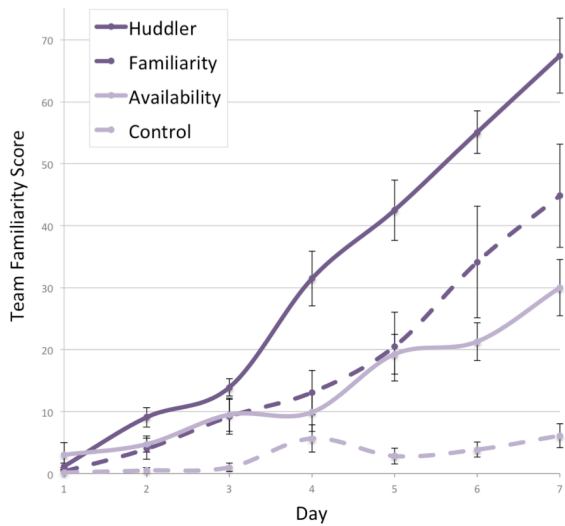

**Figure 7. Familiarity: Teams in the Huddler condition had significantly higher familiarity scores. Familiarity scores are the number of past collaborations weighted by mutual ratings.**

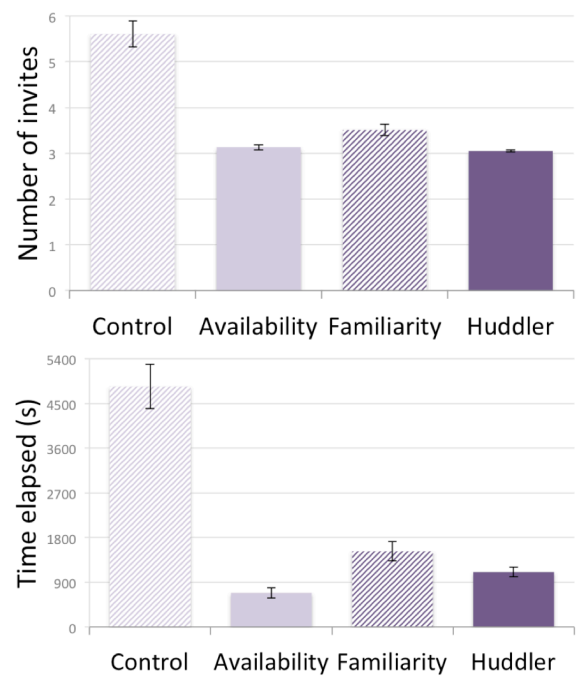

**Figure 8. Availability: Conditions that factored personal availability for each worker were faster to form teams and sent out fewer invitations.**

Towards the end of our experiment, this difference grew. By day 7, Huddler teams had the highest familiarity score (M=67.4, SD=26.3), followed by the familiarity condition (M=44.8, SD=28.8), the availability condition (M=30, SD=15), and the control condition (M=6.1, SD=5.5). We performed a two way ANOVA using team familiarity score as dependent variable and familiarity and availability factors as independent variables. The ANOVA found a significant effect of familiarity ($F(1, 40) = 28.9$, $p < .001$) and of availability $F(1, 40) = 11.2$, $p < .01$), and no interaction effect ($F(1, 40) = 0.04$, $p > .05$).

*The high availability conditions generally formed faster teams, with fewer invitations (Figure 8)*. We measure wait time as the time elapsed between when the first member applies to the task, and when all members have accepted the task and it has started. The control condition took on average 80 minutes to form teams (SD=7.3). The Availability condition had the lowest wait time, which was 11.4 minutes on average (SD=1.7). Huddler teams waited on average 18.4 minutes (SD=1.6) for their team to form.

We performed a two way ANOVA using wait time as dependent variable and familiarity and availability as independent variables. The ANOVA found a significant effect of familiarity ($F(1, 244) = 29.5$, $p < .01$) and of availability ($F(1, 244) = 88.1$, $p < .01$), but there was a significant interaction effect ($F(1, 244) = 91.3$, $p < .01$). Therefore, we explored the nature of the interaction with post-hoc tests. All pairwise comparisons were significant ($p < .05$), except for the effect of familiarity on high availability conditions (Huddler vs. availability) and the effect of availability on high familiarity conditions (Huddler vs. familiarity).

We also consider the number of invites sent out for each team until the team was formed. The minimum number of invites required is 3. In the Huddler condition, most invites were accepted and each team required 3.04 invites on average (SD=0.21). The control condition was much worse; teams required on average 5.6 invites (SD=1.84).

We performed a two way ANOVA using the number of invitations as dependent variable and familiarity and availability as independent variables. We found a significant effect of familiarity ($F(1, 244) = 52.7$, $p < .01$) and of availability ($F(1, 244) = 102.6$, $p < .01$), but there was a significant interaction effect ($F(1, 244) = 72.8$, $p < .01$). We explore the nature of the interaction with post-hoc tests. All pairwise comparisons were significant ($p < .05$) except for between the availability condition and either the familiarity condition or the Huddler condition (Huddler vs. availability, and familiarity vs. availability).

*The availability factor resulted in fewer people completing tasks*. One of Huddler's goals is to enable teams to form as quickly as possible without losing the benefits of familiarity. However, when Huddler was attending to availability, it became risk averse and preferred workers who it knew would respond quickly. As a result, the Huddler and availability conditions tended to stick with the first group of people who they identified as highly available and continued to invite them to team tasks. Many others were never invited.

Conditions that had high availability became strongly clustered. In the high availability conditions (Huddler and availability), the top 10% of workers completed 69% of tasks. Low availability conditions (familiarity and control) were more flat: the top 10% of workers completed 49% of tasks. To enable fairer access to tasks for all workers and to enable better teams to form, future systems should balance availability with engaging newcomers.

**Qualitative feedback**
We complement our analytical results with interviews and surveys where we asked crowd workers to reflect on their experience working with teams.

**Method**
We contacted 12 participants (3 people from each of the 4 conditions), prioritizing more active Huddler users, and requested interviews. Of those participants, 10 responded and we engaged in semi-structured interviews with them using Huddler's chat. We focused the interviews on what mechanisms they used to complete the work, scenarios where communication had worked well or was frustrating, and how the experience of working with familiar teams differed from new teams.

Two members of our research team read the interviews and extracted common themes. To ensure that the themes that we had identified were in fact common experiences, we then sent out a survey to all Huddler users who had completed tasks (N=74) to evaluate our themes.

**Results**
We extracted the following common themes from our interviews with Huddler workers.

*Familiar teams learned each other's strengths and used them more effectively*. In line with prior literature [33, 45], participants also mentioned performance benefits from working with familiar teams, one said:

> *"I think having teams was great because you started to learn about what people's strengths were, and we were able to use those in order to come up with the best ideas and processes for the tasks."*

Other participants mentioned the benefit of knowing *"what to expect"* from a familiar teammate and that after completing a number of tasks together *"we got into a groove and knew exactly what we were doing."* We validated these results with our survey and found that 64.8% agreed that it was beneficial to work with the same people on multiple tasks because they understood and were able to take advantage of each other's strengths.

*Teams experienced psychological safety and acceptance.* One of the most important conditions that enable teams is a shared belief that the team is a place where members feel accepted and respected. In such teams, members feel safe to take risks and generate new ideas without fear of being judged. All of these characteristics contribute to a team's psychological safety [14]. It is not clear whether anonymous, online teams would experience high psychological safety. Prior work has found that such groups might instead become breeding grounds for anxiety that would counter the benefits of working with a team [10].

In our interviews, participants reported team dynamics that suggest high levels of psychological safety, for instance:

> "I definitely felt like everyone appreciated my contributions and were at least pleased to see me in their group. I get that sense because others would often praise me and/or my inputs, make changes I suggested, and use my contributions"

To learn more, we included two related questions drawn from measures of psychological safety [14] in our survey. We found that 85.1% of workers agreed that they felt comfortable enough to contribute their own opinion when working on a team, and 81.1% agreed that they felt like everyone appreciated their contributions and skills. These provide some evidence that teams on Huddler may experience psychological safety. The knowledge that their peers will review them and they may work together again in the future may have contributed to more psychologically safe teams. Future research will explore this factor in more depth and analyze the circumstances that contribute to it.

*Teams experienced varying levels of cooperation and conflict.* Most workers who we talked to had not experienced any trouble achieving agreement with their teams, however some had other experiences:

> "For most teams, we agreed with no trouble at all. However, there were a couple of teams which didn't agree and it was because there was a team member or two who only wanted their own ideas used […] on one occasion the one team member insisting his be used alone never would back down. So we simply let him turn it in that way and then rated him at the bottom of the scale."

In our survey 18.9% of workers reported that they had at least one experience with an uncooperative teammate. Using Huddler, workers can rate uncooperative teammates low and as a result reduce their chances of working with those members again. As crowd work advances from structured workflows and individualized tasks to engage in more complex and interrelated collaborations, conflicts will be inevitable. Future work can design new mechanisms to help resolve such conflicts [53].

*One of the biggest challenges for workers was organizing their team.* When we asked about the disadvantages of working on teams using Huddler, participants consistently mentioned the difficulty of getting organized especially at the beginning. One participant said:

> "It can be tricky getting everyone in the right place at the right time. And I tried to keep things moving along efficiently in my teams, but I heard others would sometimes stall in indecisiveness for much longer."

However, many participants also mentioned that organization became easier as they completed more tasks with the same people:

> "*For my last couple of tasks as it was a quite familiar team by that point so we could relax and just have fun with the creative ideas*"

In our follow up survey we found that 32.5% of participants felt frustration from having to wait for their teammates to join so that they could start the task. However, of these participants, 72% were in the low availability condition. Therefore Huddler's availability algorithm can help reduce the difficulties of organization.

**DISCUSSION AND FUTURE WORK**
In this paper we have explored the possibility of enabling crowd workers to form familiar teams who work effectively together. We identified variable availability of team members as a major challenge for these teams and presented technical solutions to address this challenge. In a field experiment, modeling familiarity and availability were both necessary to form highly familiar teams quickly.

One challenge that Huddler teams faced was a team member accepting a task, but not showing up to work on it. Overall, 11 out of 280 tasks had one missing member, and 1 task had two missing members. In these cases, the team members who were present worked on the task and we did not pay the missing member. The familiarity and availability factors reduced this problem to some extent: 6 out of the 11 tasks that had one missing member and the task that had two missing members were all in the control condition. Future systems may give workers a way to request a backup member on-demand if they are ready to start working but one member is not present.

Occasionally, a team member would cease cooperating with their team and submit the work pre-maturely. However, this was rare and out of the 280 tasks that crowd workers completed on Huddler, only 3 contacted us and requested to change their submitted work. One mechanism that Huddler uses to prevent this problem is gathering peer assessments that it uses to form future teams. Therefore, workers can signal uncooperative members to the system. However, Huddler tasks were relatively short and took 10–20 minutes to complete: uncooperative team members may be more costly for projects that last days or weeks. Such projects may benefit from team-building pre-tasks and ongoing evaluations.

A limitation of our study is that we tested our hypothesis with one type of task and with a short time limit. We rely on prior research that has found the same positive effects of familiarity in long-term collaborations in addition to short-term tasks [26]. Future research will investigate the effects of the length of collaboration and type of task on familiarity in crowd teams.

A limitation of Huddler's algorithm is its simplified model of availability. Huddler assumes the same availability regardless of time of day or week, but crowd workers have varying schedules. Early in our study, workers requested functionality to put their Huddler account on "away" mode so that they would not receive invitations on days that they were not working on AMT, since ignoring those invitations would jeopardize their availability metric on the platform. We implemented this feature. However, the availability model still has room for improvement. For instance, each person's availability will be different in different times of the day and in different time zones. To prevent this issue, during our study we posted all tasks at the same time in the morning and limited our pool to US workers. Future work will design more sophisticated models of availability.

Another limitation of Huddler's algorithm is that it does not pay attention to a worker's bandwidth and invites workers in a greedy fashion. For instance, a popular worker in the familiarity condition may receive many invites at the same time. A more sophisticated algorithm may plan out tasks better or put a limit on the number of invites that a worker can receive. Crowd workers can also manage some of this coordination. For instance, a worker asked us to add a feature to Huddler where they could easily compare their invitations and choose which ones to accept.

With Huddler, we have aimed to maximize familiarity between team members. However, there are also benefits to working with new people. While familiar teams enjoy coordination benefits and have higher productivity, prior research has found that adding new members to already established teams can promote self-reflection on team processes and enhance creativity [40, 47]. When workers build new relationships, they transfer valuable information within the organization [2]. However, too much turnover limits organization's ability to learn [7, 12]. Evidence suggests that there is a U-shaped relationship between turnover and innovation, and organizations perform best when they have a moderate amount of turnover within work teams [19, 36]. In other words, newcomers do increase creativity and stimulate innovation, but more than a certain point they disrupt the innovation process [47]. Future work will investigate these benefits and tradeoffs for crowd teams. For instance, what is the efficient number of strong connections for a crowd worker to establish? When is it best to add a new member to an established team? While traditional organizations are limited in the number of potential collaborations between employees and the overhead costs of moving people around, crowd work systems have many more opportunities both to research these mechanisms and to utilize them to support more effective teamwork.

## CONCLUSION
To date, crowdsourcing has shared many of the characteristics of traditional models of organizing work: assembly lines emphasized workflows, roles, and top-down control. These models achieved efficiency by relying on certainty, predictability, and objective measurements of individual performance. We envision a future in which crowd work is instead characterized by complex networks of skilled workers who build long-term connections and work together in stable teams that improve iteratively. With Huddler we have taken a first step towards enabling such collaborations.

## ACKNOWLEDGMENTS

We thank the crowd workers on AMT who participated in this project. This work was supported by a National Science Foundation award IIS-1351131, an Office of Naval Research grant N00014-16-1-2894, and a Stanford Graduate Fellowship.